\documentclass[aps, prl, english, notitlepage, citeautoscript, superscriptaddress, reprint]{revtex4-2}

\usepackage{xr}
\usepackage{amsmath,bm}
\usepackage{amsfonts, amssymb,amsxtra}
\usepackage[]{graphicx}
\usepackage{grffile}
\usepackage{amsfonts}
\usepackage{framed}
\usepackage{bbm}
\usepackage{braket}
\usepackage{xcolor}
\usepackage{LatexCommands}
\usepackage{multirow}
\usepackage[colorlinks=true]{hyperref}
\usepackage{amsmath,bm}
\hypersetup{
    unicode=false,          % non-Latin characters
    pdftoolbar=true,        % show Acrobat
    pdfmenubar=true,        % show Acrobat
    pdffitwindow=false,     % window fit to page when opened
    pdfstartview={FitH},    % fits the width of the page to the window
    pdftitle={},    % title
    pdfauthor={},     % author
    pdfsubject={},   % subject of the document
    pdfcreator={},   % creator of the document
    pdfproducer={}, % producer of the document
    pdfkeywords={} {} {}, % list of keywords
    pdfnewwindow=true,      % links in new window
    colorlinks=true,       % false: boxed links; true: colored links
    linkcolor=magenta, %red,          % color of internal links (change box color with linkbordercolor)
    citecolor=blue,        % color of links to bibliography
    filecolor=magenta,      % color of file links
    urlcolor=blue           % color of external links
}

\usepackage{pdfpages}
\makeatletter
\AtBeginDocument{\let\LS@rot\@undefined}
\makeatother

\begin{document}
\title{Bicircular light tuning of magnetic symmetry and topology in Dirac semimetal Cd$_3$As$_2$}

\author{Tha\'{i}s V. Trevisan}
\affiliation{Ames Laboratory, Ames, Iowa 50011, USA}
\affiliation{Department of Physics and Astronomy, Iowa State University, Ames, Iowa 50011, USA}

\author{Pablo Villar Arribi}
\affiliation{Materials Science Division, Argonne National Laboratory, Lemont, Illinois 60439, USA}

\author{Olle Heinonen}
\affiliation{Materials Science Division, Argonne National Laboratory, Lemont, Illinois 60439, USA}

\author{Robert-Jan Slager}
\affiliation{TCM Group, Cavendish Laboratory, University of Cambridge, Cambridge CB3 0HE, United Kingdom}
\affiliation{Department  of  Physics,  Harvard  University,  Cambridge  MA  02138,  USA}

\author{Peter P. Orth}
\email{porth@iastate.edu}
\affiliation{Ames Laboratory, Ames, Iowa 50011, USA}
\affiliation{Department of Physics and Astronomy, Iowa State University, Ames, Iowa 50011, USA}

\begin{abstract}
We show that Floquet engineering using bicircular light (BCL) is a versatile way to control magnetic symmetries and topology in materials. The electric field of BCL, which is a superposition of two circularly polarized light waves with frequencies that are integer multiples of each other, traces out a rose pattern in the polarization plane that can be chosen to break selective symmetries, including spatial inversion. Using a realistic low-energy model, we theoretically demonstrate that the three-dimensional Dirac semimetal Cd$_3$As$_2$ is a promising platform for BCL Floquet engineering. Without strain, BCL irradiation induces a transition to a non-centrosymmetric magnetic Weyl semimetal phase with tunable energy separation between the Weyl nodes. In the presence of strain, we predict the emergence of a magnetic topological crystalline insulator with exotic unpinned surface Dirac states that are protected by a combination of twofold rotation and time-reversal $(2')$ and can be controlled by light. 
\end{abstract}
\date{\today}
\maketitle

\emph{Introduction.--} Recent years have seen a surge of interest in symmetry-protected topological (SPT) phases with unique electronic properties arising from a nontrivial topology of the bulk band structure. Examples are topological (crystalline) insulators~\cite{hasanColloquiumTopologicalInsulators2010, QiZhang_rmp, Fu_TCI_2011, Clas2} with dissipationless metallic surface states, magnetic axion insulators with large magnetoelectric couplings~\cite{Moore_Axi, Essin_Axi, Qi_Axi}, and three-dimensional (3D) Dirac and Weyl semimetals (SM) with exotic Fermi arc surface states ~\cite{armitageWeylDiracSemimetals2018}.
The connection between symmetry and band topology is made explicit in classification schemes based on topological invariants like Chern and winding numbers~\cite{schnyderClassificationTopologicalInsulators2008}, which depend on time reversal (TR), particle-hole and chiral symmetry alone. Crystalline materials allow for a much more refined classification due to unitary symmetries using band compatibility relations~\cite{kruthoffTopologicalClassificationCrystalline2017,bradlynTopologicalQuantumChemistry2017, Bouhon_comp, poSymmetrybasedIndicatorsBand2017}, which effectively generalize the parity indicator of Fu and Kane~\cite{fuTopologicalInsulatorsInversion2007}.
A particularly rich set of SPT phases arises from magnetic symmetries, which combine TR with spatial symmetries, leading to 1651 magnetic space groups (MSGs). These were recently addressed within the context of symmetry indicators~\cite{watanabeStructureTopologyBand2018} and elementary band representations (EBRs)~\cite{elcoroMagneticTopologicalQuantum2020, Bouhon_magneticEBR}.

The relation between symmetry and topology not only facilitates the search for new material realizations of SPT phases, but can also be employed to actively tune between different topological states by applying symmetry-breaking perturbations. Among them, irradiation with strong coherent light is particularly attractive since it allows for a \emph{dynamic manipulation} on ultrafast time scales~\cite{okaPhotovoltaicHallEffect2009,kitagawaTransportPropertiesNonequilibrium2011,hubenerCreatingStableFloquet2017,okaFloquetEngineeringQuantum2019,rudnerBandStructureEngineering2020, weberUltrafastInvestigationControl2021}. Experimentally, light-control of symmetry has been demonstrated via photocurrent generation~\cite{sirica2020photocurrentdriven}, resonant and coherent phonon excitation~\cite{forstNonlinearPhononicsUltrafast2011,  sieUltrafastSymmetrySwitch2019,disaPolarizingAntiferromagnetOptical2020,zhangLightInducedSubpicosecondLattice2019,luoLightinducedPhononicSymmetry2021}, and direct Floquet-dressing of electronic states using circularly polarized light (CL) ~\cite{wangObservationFloquetBlochStates2013, mciverLightinducedAnomalousHall2020}. It was shown that TR breaking via CL irradiation induces a quantum anomalous Hall state in topological insulator (TI) surface states~\cite{wangObservationFloquetBlochStates2013} and in graphene~\cite{satoMicroscopicTheoryLightinduced2019, mciverLightinducedAnomalousHall2020}.

Here, we show that bicircular light (BCL), which consists of a superposition of two CL waves with integer frequency ratio, offers an even greater tunability. In addition to TR breaking, BCL allows to selectively break spatial symmetries \emph{including inversion} ($\mathcal{I}$), which allows the  realization of non-centrosymmetric (e.g., chiral and polar) magnetic topological phases with unique properties. In two dimensions, the effect of BCL on graphene was recently discussed theoretically in Ref.~\onlinecite{nagDynamicalSynchronizationTransition2019}. We here explore the effect of BCL in 3D and consider both semimetallic and insulating phases. We show that BCL offers great flexibility via changing parameters such as the frequency ratio, the light direction and the relative phase between the two CL components. This allows to impose a particular magnetic space group (MSG) symmetry from one the subgroups of the crystal MSG with immediate consequences on band topology. 

Focusing on a realistic model for 3D Dirac SM Cd$_3$As$_2$, we show that this material is a promising platform for BCL tuning of magnetic topological phases. This can be realized in Cd$_3$As$_2$ because the low-energy physics is determined by narrow As-4$p$ states near the Fermi energy, and other trivial bands states are well separated in energy~\cite{WangPRB2013}, especially near the $\Gamma$ point. 
First, similar to CL, BCL splits the bulk Dirac nodes and drives a transition to a Weyl SM. Moreover, since both TR and $\mathcal{I}$ symmetries are broken by BCL, partner Weyl nodes of opposite chirality can now be separated both in momentum and energy~\cite{Zyuzin2012}, leading to a nonzero gyrotropic magnetic response~\cite{armitageWeylDiracSemimetals2018,zhongGyrotropicMagneticEffect2016,maChiralMagneticEffect2015}. Importantly, the energy separation of the nodal points can be dynamically tuned via the relative phase between the two light waves. In the presence of an additional time-dependent magnetic field, we predict this modulation to result in a time-dependent gyrotropic magnetic current that is a signature of the bulk Weyl nodes. Finally, the Fermi arc surface states controlling surface transport are different for BCL and CL due to different positions of the bulk Weyl nodes.

Secondly, we investigate the effect of CL and BCL on strained Cd$_3$As$_2$, which is a strong TI. We highlight two interesting scenarios: (i) a CL-induced transition to an axion insulator state protected by $\mathcal{I}$ symmetry~\cite{turnerQuantizedResponseTopology2012}, and (ii) a BCL-driven transition to an inversion-broken magnetic topological crystalline insulator protected by $2'$ symmetry ($2'$ is a combination of TR and two-fold rotation), which features exotic unpinned surface Dirac  cones~\cite{fangNewClassesThreedimensional2015}. We calculate observable signatures in the surface state properties for both cases.

\begin{figure}[tbh]
    \centering
    \includegraphics[width = \linewidth]{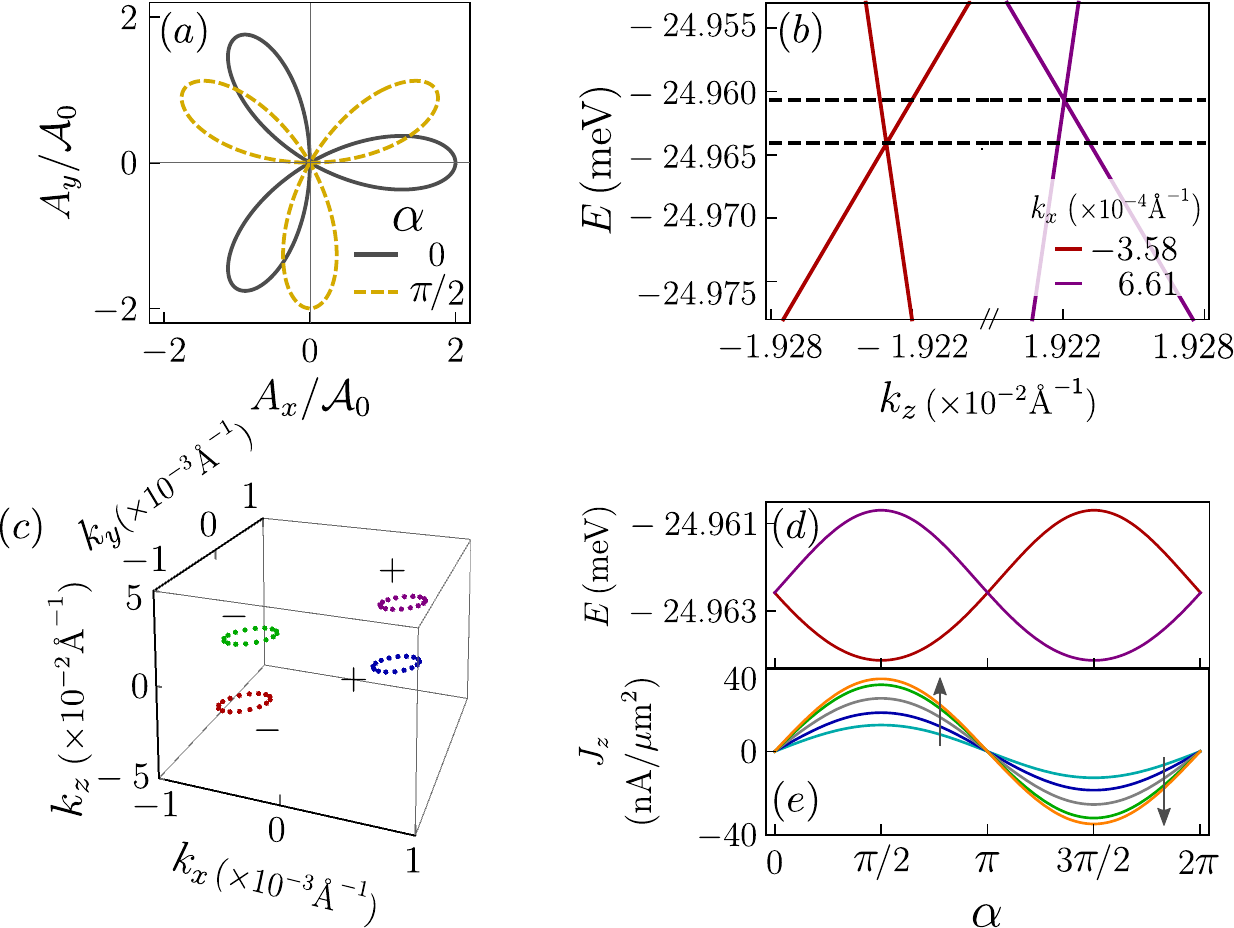}
    \caption{(a) BCL vector potential $\mathbf{A}(t)$ for $\eta=2$ and different $\alpha = 0, \frac{\pi}{2}$. $A_x, A_y$ refer to two orthogonal directions in the polarization plane. Changing $\alpha$ results in a rotation of the rose pattern. Panels (b)-(d) show location of Weyl nodes in Cd$_3$As$_2$ as a function of $\alpha$ for BCL direction normal to (112) surface. (b) Cut of the bulk bands of the effective Hamiltonian in Eq.~\eqref{eq:Heff} for fixed $k_y=k_x$ and $k_x$ as indicated. Here, we set $\omega=300$~meV, $\mathcal{A}_0=2.6\times 10^{-2}$~\AA$^{-1}$ and $\alpha=\pi/2$. A narrow energy window is shown to emphasize the energy separation of two of the Weyl nodes (see~\cite{Note1} for the full spectrum). (c) Trajectory of the Weyl nodes in momentum space as $\alpha$ evolves from $0$ to $2\pi$. The signs indicates the chirality of each node. Other parameters are identical to the ones in panel (b). Note the different scales on $k_x,k_y$ and $k_z$ axes. (d) Energy of Weyl nodes as a function of $\alpha$. (e) Gyrotropic magnetic current in response to a low-frequency oscillating magnetic field in the $\hat{z}$ direction as a function of $\alpha$. We set the amplitude of the field to $3$T. Different colors refer to distinct values of $ 0.02 \text{\AA}^{-1} \leq \mathcal{A}_0\leq 0.028$\AA$^{-1}$ with increasing steps of $0.002$~\AA$^{-1}$ in the direction of the arrows.}
    \label{fig:1}
\end{figure}

\emph{BCL Floquet-Bloch theory.--} BCL corresponds to the superposition of two CL waves with opposite chirality and different frequencies that are an integer ratio $\eta$ of each other. It is described by the vector potential 
\begin{equation} 
    \mathbf{A}(t)=\mathcal{A}_0\sqrt{2}\,\text{Re}\left[e^{-i(\eta\omega t-\alpha)}\bm{\varepsilon}_{R}+e^{-i\omega t}\bm{\varepsilon}_{L}\right]\,,
    \label{eq:A}
\end{equation}
where $\mathcal{A}_0$ is the amplitude and $\bm{\varepsilon}_{L/R}$ are left (L) and right (R) CL basis vectors~\cite{Note1}. We assume uniform light illumination and neglect the photon momentum in the following as it is much smaller than typical electronic momenta. This vector potential, and the corresponding electric field, trace out a $(\eta+1)$-fold rose curve over a period $T=2\pi/\omega$, as illustrated in Fig.~\ref{fig:1} (a). The spatial dependence of $\mathbf{A}(t)$ is crucial for breaking crystal symmetries that cannot be broken by CL, most importantly spatial inversion $\mathcal{I}$. Tuning BCL parameters $\eta$ (which sets the rose pattern symmetry), $\alpha$ (which promotes a rotation of the pattern) and the light incidence direction, results in breaking of different magnetic symmetries and thus realizes different MSGs.

In the following, we focus on three-fold BCL ($\eta = 2$) and perform a detailed study of its effects on both bulk and surface electronic states in the centrosymmetric Dirac SM Cd$_3$As$_2$~\cite{Schumann2016,Ali2014} [space group $I4_1/acd$ (No. 142)], which we describe using the realistic low-energy four-band model~\cite{WangPRB2013,PabloPRB2020}
\begin{align}
\hat{h}(\mathbf{k})&=M_{\mathbf{k}}\sigma_0\tau_z+P_{\mathbf{k}}\sigma_z\tau_x-Q_{\mathbf{k}}\sigma_0\tau_y +\varepsilon_{\mathbf{k}}\sigma_0\tau_0\,.
\label{eq:Hbare}
\end{align}

\noindent Here, $\sigma_i$, $\tau_i$ are Pauli matrices in spin and orbital space, respectively, and  $M_{\mathbf{k}}$, $P_{\mathbf{k}}$, $Q_{\mathbf{k}}$, and $\varepsilon_{\mathbf{k}}$, are $\mathbf{k}$-dependent polynomials specified in the Supplemental Material~\footnote{See Supplemental Material for details about the BCL setup, $k.p$ model for Cd$_3$As$_2$, technical details on surface state calculations and supplementary figures.}. This model has doubly degenerate valence and conduction bands that cross linearly at two Dirac nodal points along $k_z$ near the $\Gamma$-point, which are protected by four-fold rotation around $z$ ($C_{4z}$). This symmetry can be explicitly broken via application of lattice strain (leading to a transition to a strong TI phase) as we discuss in detail below. 

BCL is coupled to the electronic degrees of freedom via the Peierls' substitution. Since we consider near-infrared light frequencies well above the characteristic Lifshitz energy scale ($\approx 60$~meV) of Eq.~\eqref{eq:Hbare}, we employ the standard high-frequency approximation to arrive at a time-independent effective Floquet-Bloch Hamiltonian~\cite{MikamiPRB2016,Rahav2003,Eckardt2015,bukov2015hf}. 
Up to order $1/\omega$, it reads
\begin{align}
    &\hat{h}_{\text{eff}}(\mathbf{k})=\hat{h}_0(\mathbf{k})+\hat{m}(\mathbf{k}) \text{ ,}\label{eq:Heff}\\
    & \hat{m}(\mathbf{k})=\frac{1}{\omega}\sum\limits_{n=1}^{\infty}\frac{1}{n}\left[\hat{h}_n(\mathbf{k}),\hat{h}_{-n}(\mathbf{k})\right]\text{ ,}\label{eq:mass}
\end{align}
\noindent where $\hat{h}_n(\mathbf{k})=(1/T)\int_{0}^{T}dt\  e^{-in\omega t}\ \hat{h}[\mathbf{k}+\mathbf{A}(t)]$ is the $n$-th Fourier component of the time-dependent Hamiltonian. Exact analytical expression for $\hat{h}_{\text{eff}}$ is provided in~\cite{Note1}. Note that the polarization state of the BCL is fully encoded in Eq.~\eqref{eq:Heff}. The first term, $\hat{h}_0$, in Eq.~\eqref{eq:Heff} is the time average of $\hat{h}[\mathbf{k}+\mathbf{A}(t)]$ and merely shifts the original Dirac nodes to new locations dictated by the symmetries preserved by the light. In contrast, the second (light-induced mass) term $\hat{m}(\mathbf{k})$ contains Pauli matrix products absent in the original model, which lifts the band degeneracy and drives topological transitions. 

\emph{BCL control of SM phases.--} 
Irradiation with BCL breaks both TR and $\mathcal{I}$ symmetries and thus results in a splitting of the two Dirac nodal points into four Weyl nodes. Generically, they reside at different momenta and energies [see Fig.~\ref{fig:1}(b,c)], unless the presence of additional symmetries, which relate partner Weyl nodes with opposite chirality, forces these two to lie at the same energy. For example, for $\alpha = 0$ or $\pi$, one of the arms of the rose pattern aligns with the $x$-axis [see Fig.~\ref{fig:1}(a)] and $m_x'$ symmetry is preserved. Thus, two partner nodes connected by $m_x'$ have the same energy. 
%Note that this refers to $m_{[1\bar{1}0]'}$ symmetry for $\bfk || [112]$. 
For all other values of $\alpha$, the four nodes are separated both in energy and momentum. Thus, as $\alpha$ is varied between $0$ and $2 \pi$, the Weyl nodes trace out closed loops in energy-momentum space [see Fig.~\ref{fig:1}(c)]. The size of the loop increases with the light amplitude $\mathcal{A}_0$, which opens the possibility of light-induced braiding of the Weyl nodes. While in Cd$_3$As$_2$ we found that Weyl nodes of opposite chirality annihilate before braiding occurs, such a scenario may be possible in other models with larger Lifshitz energy or Weyl nodes in different bulk gaps. Interestingly, braiding Weyl nodes in different gaps can change their chirality, ensuring that nodes with the same chirality reside in each gap~\cite{BJY_nielsen, Eulerdrive, bouhon2019nonabelian,Wu1273, jiang2021observation2021}. As a consequence of this obstruction to annihilation of the Weyl nodes, a new invariant known as the Euler class, emerges, which is yet to be fully explored.

\begin{figure}[t!]
    \centering
    \includegraphics[width=0.99\linewidth]{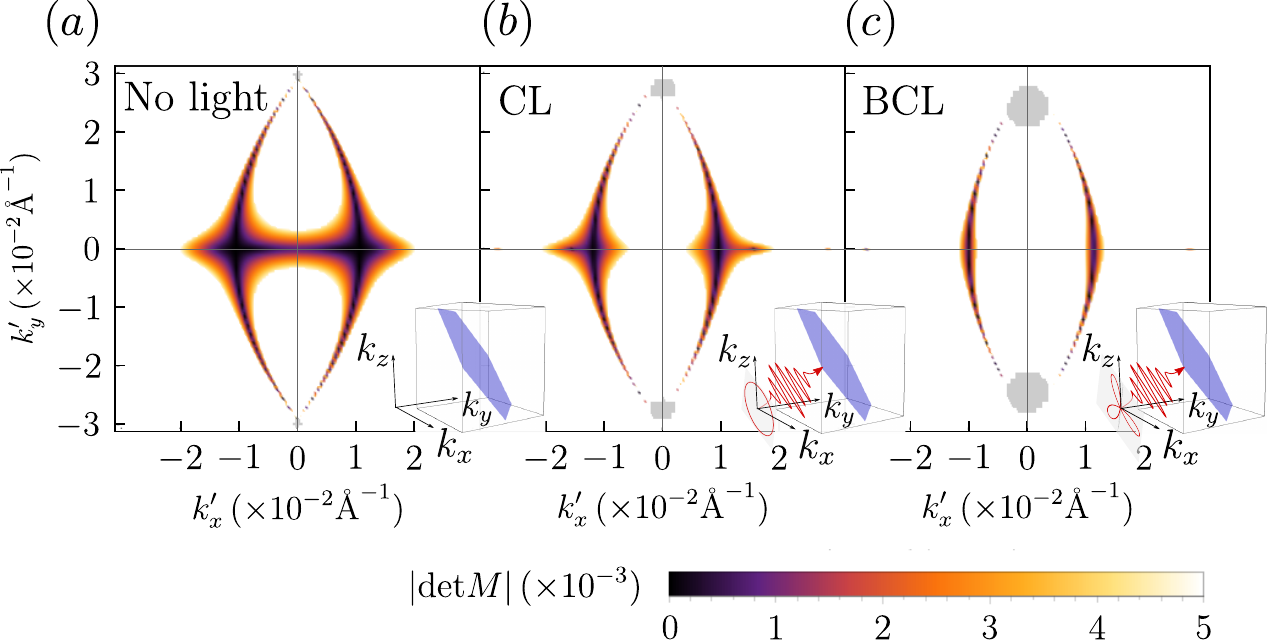}
    \caption{Fermi arc surface states at energy $E = -30$~meV on the (112) surface of Cd$_3$As$_2$ (shown in blue in the insets). $k'_{x}$ and $k'_{y}$ are components of the surface momentum along directions $[1\bar{1}0]$ and $[11\bar{1}]$, respectively. The gray regions correspond to the projection of bulk states onto the surface. Different panels show surface states: (a) in the absence of light, (b) for CL with light direction normal to (112), (c) for BCL with light direction normal to (112). In (b) and (c) we set $\omega=300$~meV, $\mathcal{A}_0=2\times 10^{-2}$\AA$^{-1}$ and $\alpha=0$. The condition $\text{det }M=0$ reflects our choice of boundary condition (see Ref.~\onlinecite{Note1}). Note that the results in panel (c) only weakly depend on the value of $\alpha$.}
    \label{fig:2}
\end{figure}

The dynamical manipulation of the energy separation between the Weyl nodes (via changing the parameter $\alpha$ as a function of time) has another interesting consequence, if one applies an additional slowly oscillating magnetic field $\bfbb(t)$ (e.g., with frequency in the megahertz range). Due to the gyrotropic magnetic effect (GME)~\cite{zhongGyrotropicMagneticEffect2016,maChiralMagneticEffect2015}, a current $\bfjj \propto \bfbb$ is generated that is proportional to the energy separation between partner Weyl nodes: $J=(e^2/3h^2) (E_{+}-E_{-})B$. As shown in Fig.~\ref{fig:1}(e), the resulting GME current is proportional to the BCL intensity and lies in the experimentally accessible nA/$\mu$m$^2$ range for $E$-field amplitudes $E_0 \approx 10^{7}$~V/m and $B\approx 3$~T. To compute $J$, we have set the chemical potential equal to the energy of one pair of Weyl nodes at $\alpha = 0$.
We propose that dynamically tuning $\alpha(t)$ on a frequency scale different from that of the magnetic field (e.g., in the kilohertz range) modulates the gyrotropic current on the same frequency, which provides a unique signature of the presence of bulk Weyl nodes. 

Finally, light irradiation also impacts the shape, curvature and position of the Fermi arc surface states in Cd$_3$As$_2$, which allows to manipulate surface transport properties~\cite{Resta2018}. We calculate the surface states of Eq.~\eqref{eq:Heff} for a half-infinite slab geometry~\cite{Note1}. Figure~\ref{fig:2} shows surface states on the experimentally accessible (112) surface~\cite{Schumann2016} at fixed energy, both in the absence of light and in the presence of CL and BCL. The Fermi arcs connect projections of the nodal points onto the surface, which are different for CL and BCL~\cite{Note1}.

\begin{figure}[t!]
    \centering
    \includegraphics[width=0.9\linewidth]{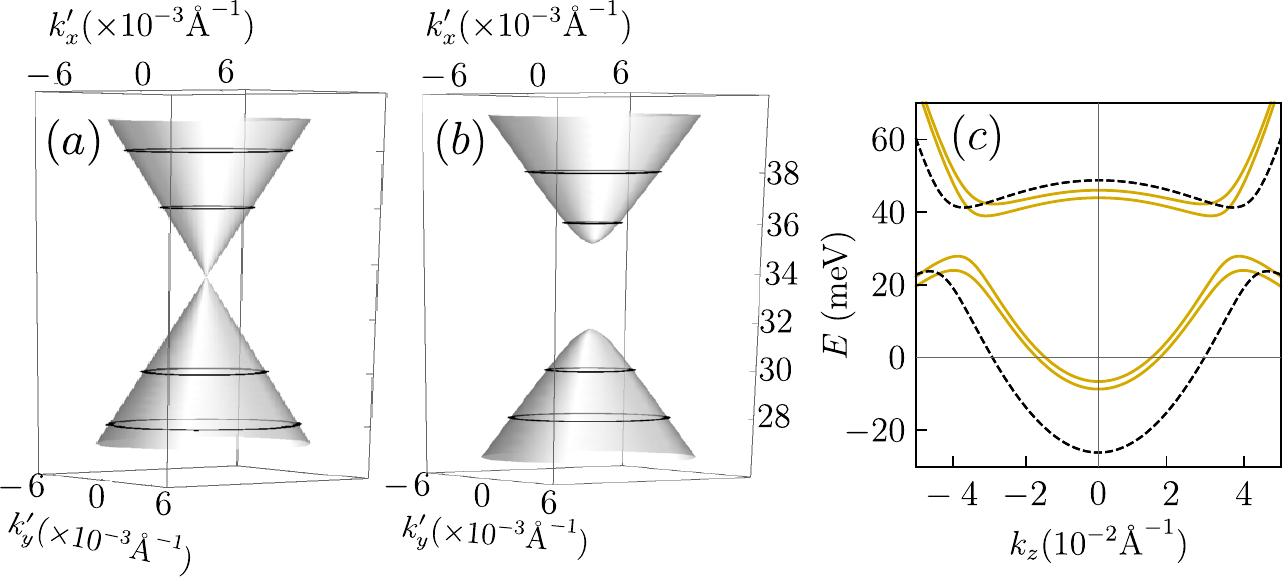}
    \caption{Surface states in the (001) surface of Cd$_3$As$_2$ under $B_{1g}$ strain (a) in the absence of light and (b) in the presence of CL with $\omega=400$~meV and $\mathcal{A}_0=2.4\times10^{-2}$\AA$^{-1}$, incident normal to (112), as in the inset of Fig.~\ref{fig:2}(b). Here, $k'_{x}$ ($k'_{y}$) points along the $[100]$ ($[010]$) direction. (c) Cut of the bulk bands of strained Cd$_3$As$_2$ for fixed $k_x=k_y=0$ in absence (dashed) and presence (solid) of CL light with the same parameters as in panel (b).}
    \label{fig:3}
\end{figure}

\emph{BCL control of insulating phases.--} Under the influence of lattice strain that breaks $C_{4z}$ symmetry, Cd$_3$As$_2$ becomes a strong TI with gapless Dirac surface states on every crystal surface. In the following, we consider experimentally realistic types of strain: first, $B_{1g}$-symmetric strain that breaks $C_{4z}$ and the two-fold rotations along the diagonals, $C_{2,[110]}$ and $C_{2,[1\bar{1}0]}$~\cite{PabloPRB2020,Schumann2016}. We also consider $B_{2g}$-symmetric strain that breaks $C_{4z}$ and the two-fold rotations along the crystallographic axes, $C_{2,[100]}$ and $C_{2,[010]}$. Within the low-energy model in Eq.~\eqref{eq:Hbare}, $B_{1g}$ and $B_{2g}$ strains correspond to terms of the form $B(\mathbf{k})\sigma_x\tau_x$ and $B(\mathbf{k})\sigma_y\tau_x$, respectively, where $B(\mathbf{k})\propto k_{z}$. Additional terms renormalizing the parameters in Eq.~\eqref{eq:Hbare} are allowed by symmetry, but vanish along $k_z$ and thus do not contribute to the size of the bulk gap (for details see~\cite{Note1}). As we show next, irradiation of strained Cd$_3$As$_2$ with CL or BCL opens up an interesting possibility of inducing exotic topological states protected by magnetic symmetries. Specifically, we predict a CL-induced axion insulator state~\cite{turnerQuantizedResponseTopology2012} and a BCL-induced magnetic topological crystalline insulator protected by $2'$ symmetry~\cite{fangNewClassesThreedimensional2015, Ahn_2019}.

\begin{table}[t!]
      \centering
      \begin{tabular}{c|c|c|c|c}
    \hline
       \text{Strain} & \text{Light type} & $\alpha$ & \text{MPG gen.} & MSG \\
         \hline \hline
         \multirow{4}{*}{0}& \text{No light} & $\times$ & $C_{2z}$, $C_{4z}$, $C_{2y}$, $\mathcal{I}$, $\Theta$ & $I4_1/acd1'$\\ \cline{2-5}
        & \text{CL} $\perp (112)$ & $\times$ & $\mathcal{I}$, $\Theta C_{2,[1\bar{1}0]}$ & $C2'/c'$
         \\ \cline{2-5}
          &  \multirow{2}{*}{\text{BCL}$\perp(112)$} &  $0$, $\pi$ & $\Theta M_{[1\bar{1}0]}$ & $Cc'$\\
          & & $\frac{\pi}{2}$, $\frac{3\pi}{2}$ & $\Theta C_{2,[1\bar{1}0]}$ & $C2'$\\
         \hline
         \hline
        \multirow{2}{*}{$B_{1g}$} & \text{No light} & $\times$ & $C_{2z}$, $C_{2y}$, $\mathcal{I}$, $\Theta$ & $Pcca1'$\\  \cline{2-5}
        & CL$\perp$(112) & $\times$ & $\mathcal{I}$ & $P\bar{1}$\\
          \hline
          \hline
         \multirow{4}{*}{$B_{2g}$} & \text{No light} & $\times$ & $C_{2z}$, $C_{2,[110]}$, $\mathcal{I}$, $\Theta$ & $Fddd1'$\\ 
        \cline{2-5}
         & \multirow{3}{*}{BCL$\perp\!(112)$} & $0$, $\pi$ & $\Theta M_{[1\bar{1}0]}$ & $Cc'$\\  
         & & $\frac{\pi}{2}$, $\frac{3\pi}{2}$ & $\Theta C_{2,[1\bar{1}0]}$ & $C2'$\\
         & & \text{N.A.} &  \text{1} & $P1$
         \\
         \hline
    \end{tabular}
    \caption{Magnetic symmetries of Cd$_3$As$_2$ in presence and absence of light and lattice strain. Light incidence direction is perpendicular to the (112) surface and BCL parameter $\eta = 2$ (three-fold rose pattern). The table shows the generators of the magnetic point group (MPG) preserved by the light and the resulting MSG for Cd$_3$As$_2$. $\Theta$ refers to TR symmetry and $\mathcal{I}$ to spatial inversion. N.A. stands for none of the above. The relative phase difference $\alpha$ rotates the BCL rose pattern, which controls the MSG. }
    \label{tab:symmetries}
\end{table}

Table~\ref{tab:symmetries} shows that the combination of $B_{1g}$ strain and CL irradiation along the (112) normal removes all symmetries except $\mathcal{I}$. Since the bulk gap remains open for not too large light intensities, we can conclude that CL induces a topological transition to an axion insulator state with quantized magnetoelectric coupling and half-quantized anomalous surface Hall conductivity~\cite{Moore_Axi}. As $\mathcal{I}$ is naturally broken at the surface, the surface states acquire a gap as shown in Fig.~\ref{fig:3} for the particular case of the $(001)$ surface. 

\begin{figure}[t!]
    \centering
    \includegraphics[width=\linewidth]{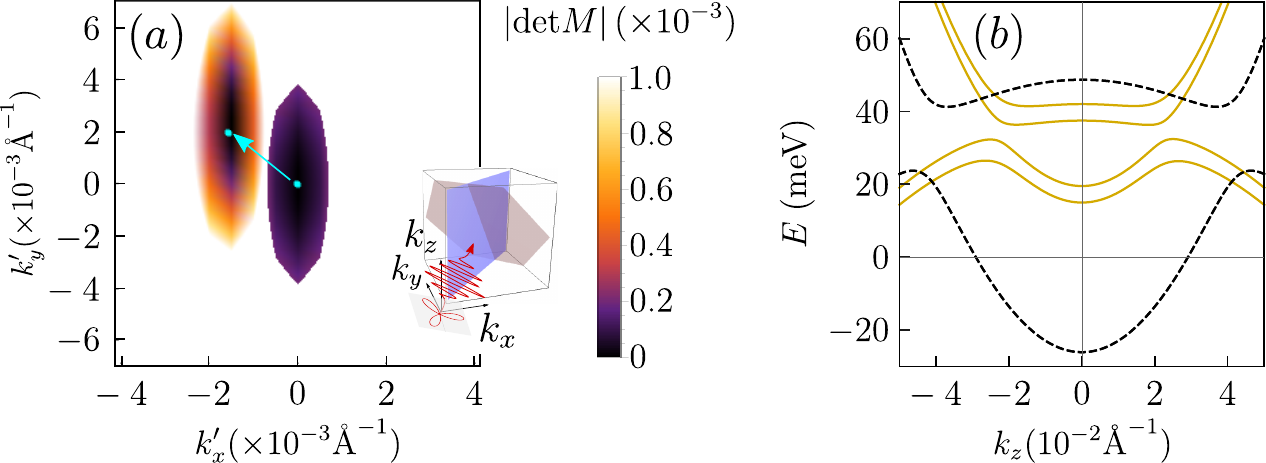}
    \caption{(a) Evolution of gapless surface states in the ($1\bar{1}0$) surface of Cd$_3$As$_2$ (blue plane in the inset) in presence of $B_{2g}$ strain when turning on BCL. Here, $k'_{x}$ ($k'_{y}$) are components along the $[110]$ ($[001]$) directions. The ellipse centered around $k'_x=k'_y=0$ is a fixed-energy cut of the surface state at $37$~meV close to the node in the absence of light. BCL irradiation along direction normal to $(112)$ (red plane in the inset) induces a $2'$ magnetic topological crystalline insulator with unpinned Dirac surface states. The surface state thus moves in the direction indicated by the arrow when BCL is turned on. The energy of the Dirac node changes only slightly to $36$~meV in the presence of BCL. Color shows $\text{det}M$, which has the same meaning as in Fig.~\ref{fig:2} (see also~\cite{Note1}). (b) Bulk energy bands of strained Cd$_3$As$_2$ for fixed $k_x=k_y=0$ without (dashed) and with BCL (solid). In both panels, we set $\omega=100$~meV, $\mathcal{A}_0=2.6\times 10^{-2}$\AA$^{-1}$, $\alpha=\pi/2$.}
    \label{fig:4}
\end{figure}

An even more intriguing situation arises in the presence of $B_{2g}$ strain and BCL irradiation along $(112)$ normal. Choosing $\alpha$ as either $\frac{\pi}{2}$ or $\frac{3\pi}{2}$ removes all symmetries except the combination of a two-fold rotation around $[1\bar{1}0]$ and time reversal: $\Theta C_{2,[1\bar{1}0]}\equiv 2'_{[1\bar{1}0]}$. Since a $2'$ operation reverses an odd number of spacetime coordinates (like TR and $\mathcal{I}$), the magnetoelectric coupling is still quantized~\cite{VanderbiltBook}. Again, for light intensities that leave the bulk gap open, continuity ensures that BCL induces a topologically nontrivial axion insulator state. Here, however, not all surfaces are gapped. Instead, $(1\bar{1}0)$ surfaces (whose surface normal is parallel to the $2'$ axis) host exotic gapless Dirac states that have a nodal position, which is unpinned from the surface TR invariant momenta. As shown in Fig.~\ref{fig:4}, the position of the nodal point is controlled by light parameters such as frequency and intensity~\cite{Note1}. All other surfaces are gapped and exhibit a half-quantized anomalous surface Hall conductivity. 

\emph{Conclusion.--} To conclude, we show that BCL irradiation offers wide tunability of the magnetic symmetries of a material beyond the capabilities of linear or circularly polarized light. This arises from the fact that the electric field of BCL follows a rose pattern that breaks $\mathcal{I}$ symmetry (in addition to TR), leading to a non-centrosymmetric effective Floquet-Bloch Hamiltonian. We demonstrate that the Dirac SM Cd$_3$As$_2$ is a suitable material platform that undergoes various light-induced topological phase transitions. In the absence of strain, BCL leads to a non-centrosymmetric magnetic Weyl SM phase with bulk nodes separated in both energy and momentum. The dynamic manipulation of the Weyl nodes position (via changing the relative phase of BCL components) in the presence of an additional magnetic field introduces a time-dependence to the gyrotropic magnetic response, which is a unique signature of Weyl physics. Finally, combining lattice strain and BCL irradiation realizes a sought-after magnetic axion insulator phase with exotic unpinned surface states protected by $2'$ symmetry. 

\begin{acknowledgements}
We thank A.~Burkov, V.~L. Quito, S.~Stemmer and D.~Yarotski for useful discussions. This research was supported by the Center for Advancement of Topological Semimetals, an Energy Frontier Research Center funded by the U.S. Department of Energy Office of Science, Office of Basic Energy Sciences, through the Ames Laboratory under Contract No. DE-AC02-07CH11358.  R.-J.~S. acknowledges funding from the Marie Sk{\l}odowska-Curie programme under EC Grant No. 842901 and the Winton programme as well as Trinity College at the University of Cambridge. 
\end{acknowledgements}

\bibliography{Biblio_BCL}

\pagebreak
\includepdf[pages={{},1,{},2,{},3,{},4,{},5,{},6,{},7,{},8,{},9,{},10,{},11,{},12,{},13,{},14}]{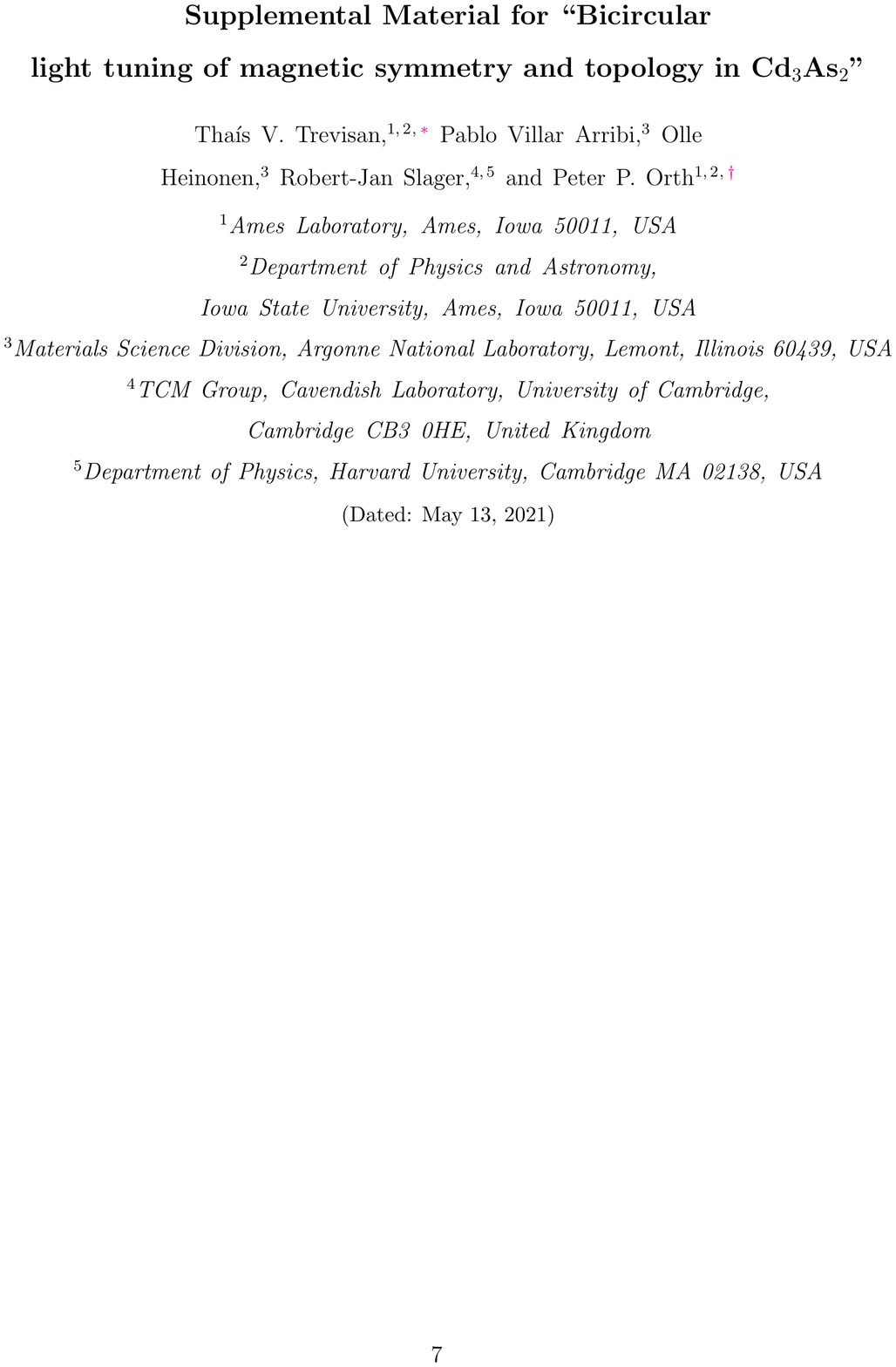}

\end{document}

\title{Supplemental Material for ``Bicircular\\ light tuning of magnetic symmetry and topology in Cd$_3$As$_2$''}

\author{Tha\'{i}s V. Trevisan}
\email{thais@iastate.edu}
\affiliation{Ames Laboratory, Ames, Iowa 50011, USA}
\affiliation{Department of Physics and Astronomy, Iowa State University, Ames, Iowa 50011, USA}

\author{Pablo Villar Arribi}
\affiliation{Materials Science Division, Argonne National Laboratory, Lemont, Illinois 60439, USA}

\author{Olle Heinonen}
%\email{heinonen@anl.gov}
\affiliation{Materials Science Division, Argonne National Laboratory, Lemont, Illinois 60439, USA}

\author{Robert-Jan Slager}
%\email[]{rjs269@cam.ac.uk}
\affiliation{TCM Group, Cavendish Laboratory, University of Cambridge, Cambridge CB3 0HE, United Kingdom}
\affiliation{Department  of  Physics,  Harvard  University,  Cambridge  MA  02138,  USA}

\author{Peter P. Orth}
\email{porth@iastate.edu}
\affiliation{Ames Laboratory, Ames, Iowa 50011, USA}
\affiliation{Department of Physics and Astronomy, Iowa State University, Ames, Iowa 50011, USA}
\date{\today}
\maketitle
\onecolumngrid
\tableofcontents
\appendix
\setcounter{page}{7}
\section{BCL setup}\label{SM_Fig1}
% Include notation for directions and surfaces
%It is also important to mention that the crystallographic axes $\hat{a}\equiv[100]$, $\hat{b}\equiv[010]$ and $\hat{c}=[001]$ are aligned with $\hat{x}$, $\hat{y}$ and $\hat{z}$, respectively. Besides, the centrosymmetric Cd$_3$As$_2$, which is our focus in this work, has a tetragonal structure with lattice parameters $a=b=12.633$\AA and $c=25.427$\AA~\cite{}.

The polarization state of the BCL field, 
\begin{equation} 
    \mathbf{A}(t)=\mathcal{A}_0\sqrt{2}\,\text{Re}\left[e^{-i(\eta\omega t-\alpha)}\,\boldsymbol{\varepsilon}_{R}+e^{-i\omega t}\boldsymbol{\varepsilon}_{L}\right]
    \label{eq:SMA}
\end{equation}
\noindent is specified by its incidence direction $\hat{q}$, frequency ratio $\eta$ between the two combined CL and their phase difference $\alpha$. The dependence on $\hat{q}$ is encoded in the vectors $\boldsymbol{\varepsilon}_{R}$ and $\boldsymbol{\varepsilon}_{L}$ that spam the light polarization plane. Here, we make this dependence more explicit. 

\begin{figure}[b!]
    \centering
    \includegraphics[width=0.85\linewidth]{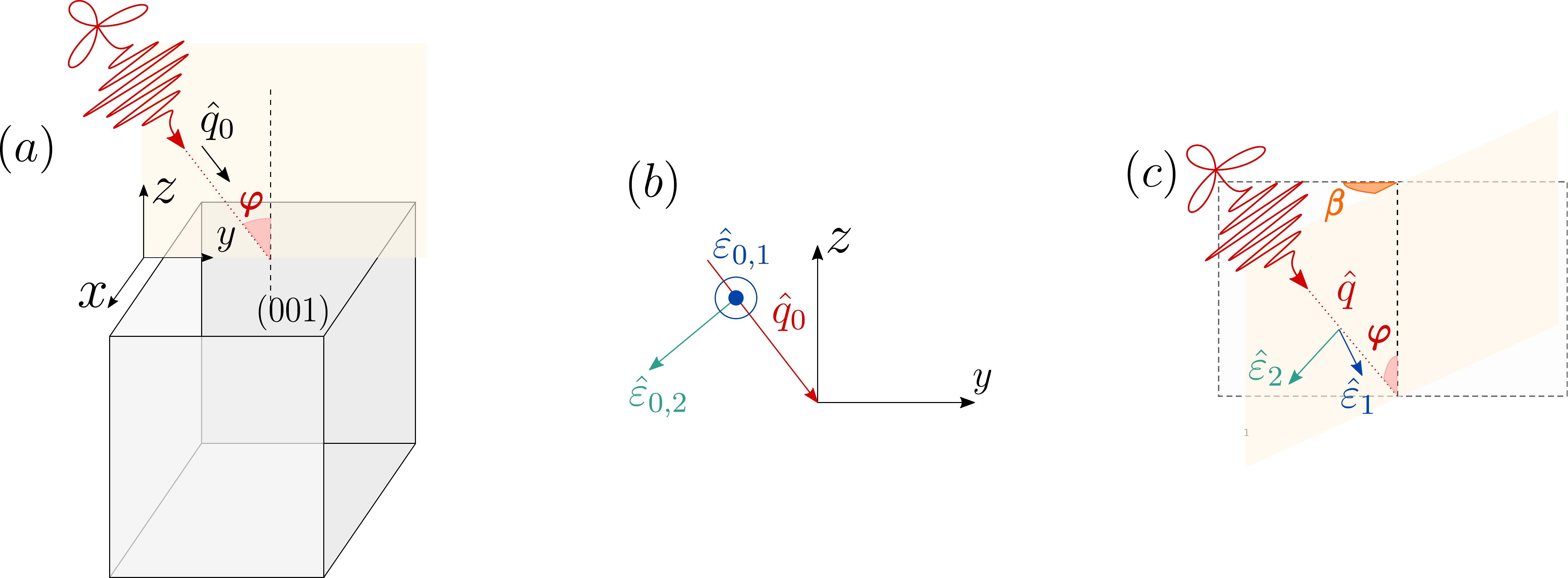}
    \caption{(a) Illustration of the tetragonal Cd$_3$As$_2$ crystal irradiated by a BCL incident with an angle $\varphi$ with respect to the normal of the (001) surface. (b) Light propagation direction $\hat{q}_0$ and vectors $\boldsymbol{\varepsilon}_1$ and $\boldsymbol{\varepsilon}_2$ defining the polarization plane for the setup shown in (a). Incidence direction and polarization vectors for a generic incidence direction.}
    \label{fig:setup}
\end{figure}

We choose $\hat{x}$, $\hat{y}$ and $\hat{z}$ to coincide with the Cd$_3$As$_2$ crystallographic axes $\hat{a}\equiv[100]$, $\hat{b}\equiv[010]$ and $\hat{c}=[001]$. A BCL is then incident with an angle $\varphi$ with respect to the normal of the (001) surface of the material, i.e. the surface perpendicular to $\hat{z}$, as illustrated in Fig.~\ref{fig:setup}(a). In this case, the incidence direction is simply $\hat{q}_0=(0,\sin\varphi,-\cos\varphi)$, which is the same incidence direction of the CLs combined in the BCL wave. Their polarization plane is spanned by the vectors $\hat{\varepsilon}_{0,1}$ and $\hat{\varepsilon}_{0,2}$. We set  $\hat{\varepsilon}_{0,1}=\hat{x}=(1,0,0)$, and $\hat{\varepsilon}_{0,2}=(0,-\cos\varphi,-\sin\varphi)$ is obtained by rotating $\hat{\varepsilon}_{0,1}$ by $\pi/2$ around $\hat{q}_0$ in order to satisfy $\hat{\varepsilon}_{0,1}\times \hat{\varepsilon}_{0,2}=\hat{q}_0$ (see Fig.~\ref{fig:setup}(b)).  

A more generic incidence direction can be straightforwardly constructed from this simple case by rotating the incident plane by an angle $\beta$ around $\hat{z}$, as illustrated in Fig.~\ref{fig:setup}(c). In this case, we have    
\begin{align}
& \hat{q}=R_{\hat{z}}(\beta)(0,\sin\varphi,-\cos\varphi)=\left(-\sin\beta\sin\varphi,\cos\beta\sin\varphi,-\cos\varphi\right) \text{ .}\\
& \hat{\varepsilon}_1=R_z(\beta)(1,0,0)=(\cos\beta,\sin\beta,0) \text{ ,}\\
& \hat{\varepsilon}_2=R_z(\beta)(0,-\cos\varphi,-\sin\varphi)=(\cos\varphi\sin\beta,-\cos\varphi\cos\beta,-\sin\varphi) \text{ .}
\end{align} 

\noindent For a light circularly polarized to the right, the polarization is specified by $\boldsymbol{\varepsilon}_{R}=\left(\hat{\varepsilon}_1-i\hat{\varepsilon}_2\right)/\sqrt{2}$, while the polarization state of a light circularly polarized to the left is given by $\boldsymbol{\varepsilon}_{L}=\left(\hat{\varepsilon}_1+i\hat{\varepsilon}_2\right)/\sqrt{2}$. Substituting $\boldsymbol{\varepsilon}_{R}$ and $\boldsymbol{\varepsilon}_{L}$ into Eq.~\eqref{eq:SMA}, we can write the Cartesian components of the BCL vector potential in a compact form:    
\begin{equation}
A_{\mu}(t)=\mathcal{A}_0\,c_{\mu}\left[\cos(\eta\omega t-\alpha)+\cos(\omega t)\right]+\mathcal{A}_0\,d_{\mu}\left[\sin(\eta\omega t-\alpha)-\sin(\omega t)\right] \text{ ,}
\label{eq:Aalpha}
\end{equation} 
\noindent where $\mu=x,y,z$. The coefficients $c_{\mu}$ and $d_{\mu}$ are parametrized by the two angles $\varphi$ and $\beta$, as shown in Table~\ref{tab:coefs}. 

Note that a BCL perpendicular to (001) or, equivalently, parallel to $\hat{z}=[001]$ is obtained by setting $\beta=0$ and $\varphi=\pi$. Since Cd$_3$As$_2$ is a tetragonal crystal, the correspondence between the indices for the surface $(hkl)$ and its normal $[uvw]$ follows~\cite{crystallography}
\begin{equation}
    \frac{u}{h}=\frac{v}{k}=\frac{w}{l}\left(\frac{c}{a}\right)^2 \text{ ,}
\end{equation}

\noindent where $a=b$ and $c$ are the lattice parameters in the $xy$ plane and in $\hat{z}$ direction, respectively. For Cd$_3$As$_2$,  $a=12.633$\AA ~and  $c=25.427$\AA~\cite{Ali2014}, but for calculations purposes we set $c=2a$. In this case, a BCL normal to (112) is parallel to the $[221]=\hat{x}+\hat{y}+\hat{z}$ direction and, therefore, obtained by setting $\beta=-\pi/4$ and $\varphi=\text{arcos}(-1/\sqrt{3})$.  

\begin{table}[b!]
\centering
\label{tab:coefs}
\begin{tabular}{|c|c|c|c|}
\hline
             & $x$                     & $y$                    & $z$            \\ \hline
$c_{\mu}$ & $\cos\beta$             & $\sin\beta$            & $0$            \\ \hline
$d_{\mu}$ & $-\sin\beta\cos\varphi$ & $\cos\beta\cos\varphi$ & $\sin\varphi$ \\ \hline
\end{tabular}
\caption{Coefficients of the Cartesian components of the BCL field.}
\end{table}
 
\section{k.p model for cadmium arsenide}

\subsection{Semimetallic phase}

Despite its complicated crystal structure, first-principle calculations~\cite{WangPRB2013} show that the low energy physics of Cd$_3$As$_2$ is dominated by the Cd 5s orbitals and the As 4p orbitals and, therefore, can be characterized by a minimal 4-band model, 
\begin{align}
&\hat{H}=\sum\limits_{\mathbf{k}}\hat{\psi}_{\mathbf{k}}^{\dag}\,\hat{h}(\mathbf{k})\,\hat{\psi}_{\mathbf{k}}^{\null} \text{ ,}\label{eq:Hk}\\
& \hat{h}(\mathbf{k})=M(\mathbf{k})\sigma_0\tau_z+P(\mathbf{k})\sigma_z\tau_x-Q(\mathbf{k})\sigma_0\tau_y+\varepsilon(\mathbf{k})\sigma_0\tau_0 \text{ .}\label{eq:hk}
\end{align}

\begin{table}[b!]
\begin{tabular}{|c|c|c|}
\hline
Parameters             & Semimetallic & Insulating \\ \hline
$M_0$ (eV) &   0.0282     &     0.0374                \\ \hline
$M_1$ (eV \AA$^{2}$) &  20.72                     &      20.36               \\ \hline
$M_2$ (eV \AA$^{2}$)&    13.32                   &     18.77                \\ \hline
$C_0$ (eV) &    -0.0475                   &   0.0113                  \\ \hline
$C_1$ (eV \AA$^{2}$)&   12.50                    &   12.05                  \\ \hline
$C_2$ (eV \AA$^{2}$)&   13.62            &   13.13                  \\ \hline
$A$ (eV\AA) &     1.116                  &   1.089                  \\ \hline
$B_1$ (eV\AA) &         0              &            0.2566         \\ \hline
\end{tabular}
\caption{Parameters for the $k.p$ model for Cd$_3$As$_2$ in both semimetallic (Dirac semimetal) and insulating (strong TI) phases. Values taken from Ref.~\cite{PabloPRB2020}. Note that in Ref.~\cite{PabloPRB2020} $M_1$ and $M_2$ are defined with a global minus sign. Here, we incorporated it directly in Eqs.~\eqref{eq:bareM} and that is why $M_1$ and $M_2$ are positive in our table.}
\label{tab:parameters_band}
\end{table}

\begin{figure}[b!]
    \centering
    \includegraphics[width=0.85\linewidth]{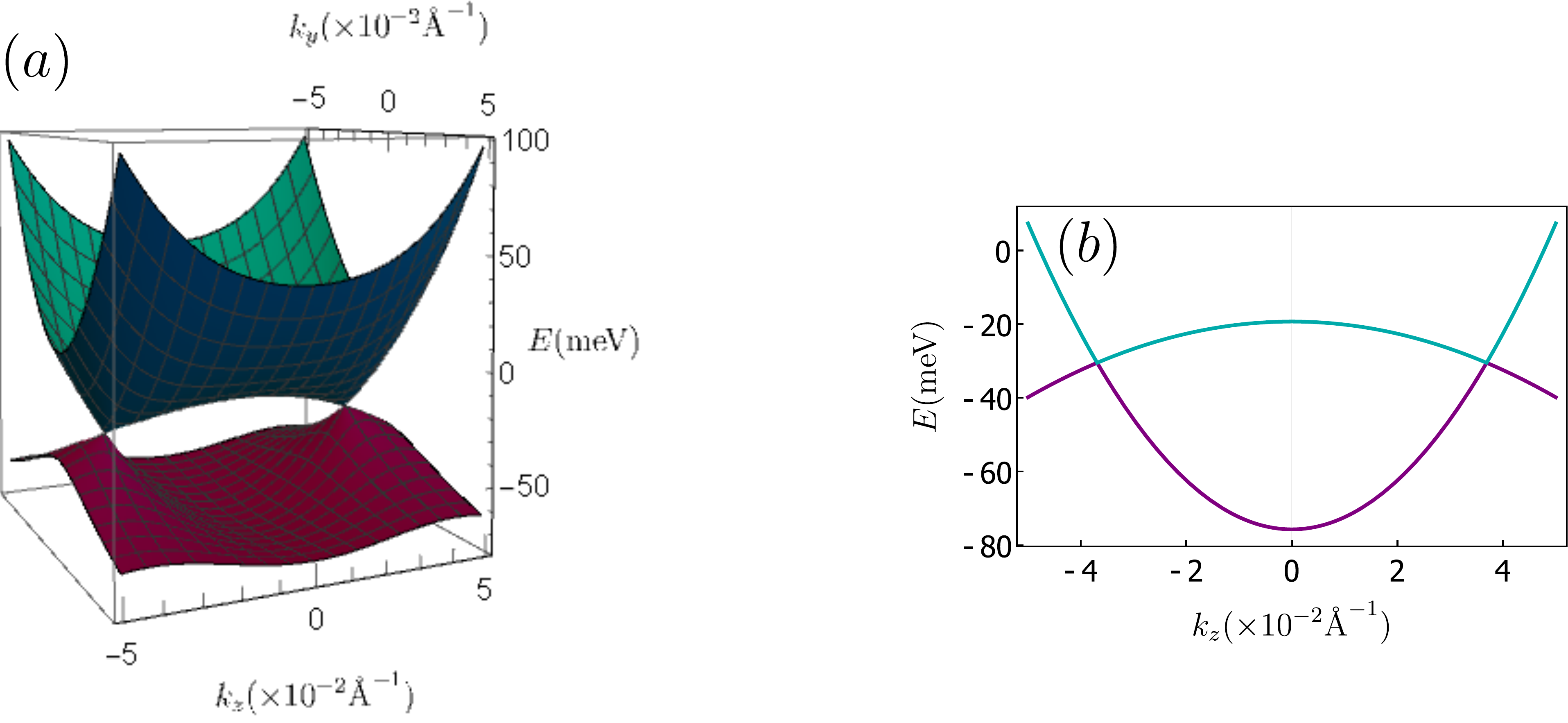}
    \caption{Bulk bands of the $k.p$ model for Cd$_3$As$_2$ in the absence of strain and light [Eq.\eqref{eq:hk}]. We set (a) $k_x=0$ and (b) $k_x=k_y=0$.}
    \label{fig:Dirac}
\end{figure}

\noindent Here,  $\hat{\psi}_{\mathbf{k}}^{\dag}\equiv (c_{\mathbf{k},\uparrow}^{\dag}d_{\mathbf{k},\uparrow}^{\dag}c_{\mathbf{k},\downarrow}^{\dag}d_{\mathbf{k},\downarrow}^{\dag})$, where $c_{\mathbf{k},\sigma}^{\dag}$ ($d_{\mathbf{k},\sigma}^{\dag}$) creates an electron with momentum $\mathbf{k}$ and spin $\sigma$ in the $5s$ ($4p$) orbital Cd (As). Besides, $\sigma_j$ ($\tau_j$), with $j=x,y,z$ are Pauli matrices in spin (orbital) space and $\sigma_0$ and $\tau_0$ are two-by-two identity matrices. The functions preceding the Pauli matrices in Eq.~\eqref{eq:hk} are polynomials of the momentum components, 
\begin{align}
& M(\mathbf{k})=M_0-M_1k_z^2-M_2\left(k_x^2+k_y^2\right)\text{ ,}\label{eq:bareM}\\
& P(\mathbf{k})=Ak_x\text{ ,}\label{eq:bareP}\\
& Q(\mathbf{k})=Ak_y \text{ ,}\label{eq:bareQ}\\
& \varepsilon(\mathbf{k})=C_0+C_1k_z^2+C_2\left(k_x^2+k_y^2\right)\text{ ,}\label{eq:bareep}
\end{align}

\noindent with coefficients selected to fit the bands obtained from DFT calculations. We used the numbers calculated in Ref.~\cite{PabloPRB2020}, which are summarized in Table~\ref{tab:parameters_band}. Such a simple $k.p$ model is able to capture the two Dirac cones in the band structure of Cd$_3$As$_2$ because they occur in the vicinity of the $\Gamma$ point ($\mathbf{k}=0$) of the Brillouin zone, as shown in Fig.~\ref{fig:Dirac}.

It is important to note that terms such as Eqs.~\eqref{eq:bareP} and \eqref{eq:bareQ} originate from $\sin (ak_x)$ and $\sin(ak_y)$, respectively, in the tight-biding model valid in the entire Brillouin zone expanded to linear order in momentum. We noticed, however, that when light is coupled to the electrons, we need to carry the expansion at least up to third order in momentum for the symmetries of the Floquet Hamiltonian to be consistent with those preserved by the light vector potential. Accordingly, we add corrections $\gamma k_x^3$ and $\gamma k_y^3$ to Eq.~\eqref{eq:bareP} and Eq.~\eqref{eq:bareQ}, respectively. Symmetry allows for the coefficients of the linear and cubic terms in Eq.~\eqref{eq:bareP} and Eq.~\eqref{eq:bareQ} to be different, and we choose $\gamma=A/6$ to ensure that the new cubic terms are only a small correction to the original expressions. Similarly, for linear terms in $k_z$ that will appear later in the presence of strain, we add a correction four times larger ($4\gamma k_z$, with $\gamma=B_1/6$ - see Eq.~\eqref{eq:B}) to reflect $c=2a$.    

\subsection{Insulating phase}
The effects of strain in the plane perpendicular to $\hat{z}$ are incorporated in the $k.p$ model by adding extra terms to Eq.~\eqref{eq:hk} that break the desired symmetries. The values of the coefficient in the polynomial functions also change (see Table~\ref{tab:parameters_band}), and the form of these symmetry-breaking terms depend on the type of strain we consider. 

Let's start with strain type $B_{1g}$. It breaks the two-fold rotations around the diagonals in the $xy$ plane $C_{2,[110]}$ and $C_{2,[1\bar{1}0]}$ (and the corresponding mirrors $M_{[110]}$ and $M_{[1\bar{1}0]}$) in addition to $C_{4z}$. All other symmetries of the unstained material remain unchanged and, as a consequence, Cd$_3$As$_2$ with $B_{1g}$ strain is characterized by the space group $Pmmn$ (No. 59). Time-reversal is also preserved in the absence of light and the corresponding magnetic space group is the grey group $Pmmn1'$ (No.59.406). The generic first-quantized Hamiltonian that preserves the symmetries of this group is given by 
\begin{equation}
    \hat{h}_{B_{1g}}(\mathbf{k})=M(\mathbf{k})\sigma_0\tau_z+B(\mathbf{k})\sigma_x\tau_x+P(\mathbf{k})\sigma_z\tau_x-Q(\mathbf{k})\sigma_0\tau_y+\varepsilon(\mathbf{k})\sigma_0\tau_0 \text{ ,}
    \label{eq:hB1g}
\end{equation}

\noindent with ~\cite{PabloPRB2020}
\begin{equation}
    B(\mathbf{k})=B_1\left(k_z+\frac{2k_z^3}{3}\right) \text{ .}
    \label{eq:B}
\end{equation}

\noindent The functions $M(\mathbf{k})$, $P(\mathbf{k})$, $Q(\mathbf{k})$ and $\varepsilon_0(\mathbf{k})$ have the same form as in the semimetallic case, but with different coefficients (see Table~\ref{tab:parameters_band}). 

Strain of type $B_{2g}$ in the $xy$ plane, on the other hand, breaks, besides $C_{4z}$, the two-fold rotations around the crystallographic axes $C_{2,x}$ and $C_{2,y}$, as well as their corresponding mirrors, but preserves $C_{2,[110]}$ and $C_{2,[1\bar{1}0]}$. The resulting space group is $Ccca$ (No. 68). The generic first-quantized $k.p$ Hamiltonian that preserves the symmetries of this group is    
\begin{equation}
    \hat{h}_{B_{2g}}(\mathbf{k})=\tilde{M}(\mathbf{k})\sigma_0\tau_z+B(\mathbf{k})\sigma_y\tau_x+\tilde{P}(\mathbf{k})\sigma_z\tau_x-\tilde{Q}(\mathbf{k})\sigma_0\tau_y+\tilde{\varepsilon}(\mathbf{k})\sigma_0\tau_0 \text{ ,}
    \label{eq:hB2g}
\end{equation}

\noindent where 
\begin{align}
    & \tilde{M}(\mathbf{k})= M(\mathbf{k})+s_{A}k_xk_y \text{ ,}\\
    & \tilde{P}(\mathbf{k})= P(\mathbf{k})+\tilde{A}k_y\left(1+\frac{k_y^2}{6}\right) \text{ ,}\\
    & \tilde{Q}(\mathbf{k})= Q(\mathbf{k})+\tilde{A}k_x\left(1+\frac{k_y^x}{6}\right) \text{ ,}\\
    & \tilde{\varepsilon}(\mathbf{k})=\varepsilon(\mathbf{k})+s_{B}k_xk_y \text{ ,}
\end{align}

\noindent and $M(\mathbf{k})$, $P(\mathbf{k})$, $Q(\mathbf{k})$, $\varepsilon(\mathbf{k})$ and $B(\mathbf{k})$ defined in Eqs.~\eqref{eq:bareM}-\eqref{eq:bareep} and Eq.~\eqref{eq:B}, respectively. For simplicity, we set, for simplicity, $s_A=s_B=\tilde{A}=B_1=0.2566$~eV\AA.   

\section{Light-induced mass terms}
Here, we show the expressions for the light-induced mass terms $\hat{m}(\mathbf{k})$ defined in the main text. Standard high-frequency expansion~\cite{MikamiPRB2016,Rahav2003,Eckardt2015,bukov2015hf} tell us that a time-independent effective Hamiltonian $\hat{H}_{eff}$ can be derived from a generic periodic Hamiltonian $\hat{H}(t)=\hat{H}(t+T)$ using 
\begin{equation}
\hat{H}_{eff}=\hat{H}_0+\frac{1}{\omega}\sum\limits_{n=1}^{\infty}\frac{1}{n}\left[\hat{H}_{n},\hat{H}_{-n}\right] +\mathcal{O}\left(\frac{1}{\omega^2}\right) \text{ ,}
\end{equation}

\noindent where 
\begin{equation}
\hat{H}_{n}=\frac{1}{T}\int\limits_{0}^{T}dt e^{-in\omega t}H(t) \text{ ,}
\end{equation}

\noindent denotes the Fourier components of $\hat{H}(t)$ (with $n=0,1,2,\cdots$). For the particular case of the many-body Hamiltonian Eq.~\eqref{eq:Hk}), the fermionic anti-commutation relation gives 
\begin{equation}
    \hat{H}_{eff}=\sum\limits_{\mathbf{k}}\hat{\psi}_{\mathbf{k}}^{\dag}\,\hat{h}_{eff}(\mathbf{k})\,\hat{\psi}_{\mathbf{k}}^{\null} \text{ ,}
\end{equation}

\noindent with 
\begin{align}
    &\hat{h}_{eff}(\mathbf{k})=\hat{h}_0(\mathbf{k})+\hat{m}(\mathbf{k})\text{ ,}\label{eq:heffSI}\\
    & \hat{m}(\mathbf{k})=\frac{1}{\omega}\sum\limits_{n=1}^{\infty}\frac{1}{n}\left[\hat{h}_n(\mathbf{k}),\hat{h}_{-n}(\mathbf{k})\right]\label{eq:mSI}\text{ ,}
\end{align}

\noindent as defined in the main text. Note that $\hat{h}_n(\mathbf{k})$ has the same form of the first-quantized Hamiltonian in Eq.~\eqref{eq:hk} (or Eq.~\eqref{eq:hB1g}, or Eq.~\eqref{eq:hB2g} depending on which phase of Cd$_3$As$_2$ we are considering), but the functions preceding the Pauli matrices are replaced by their Fourier component after minimal coupling. For instance, $M(\mathbf{k})\rightarrow M_{n}(\mathbf{k})=(1/T)\int_{0}^{T}e^{-in\omega t}M(\mathbf{k}+\mathbf{A}(t))$. 

The term proportional to $1/\omega$ in $\hat{h}_{eff}(\mathbf{k})$ [Eq.~\eqref{eq:mSI}], acts as a mass term in the location of the Dirac nodes of the unperturbed band structure and, therefore, we call it \textit{light-induced mass term}. It has nine distinct contributions as a result of the commutations of the Pauli matrices:
\begin{align}
\hat{m}(\mathbf{k})&=m_{1}(\mathbf{k})\sigma_x\tau_y+m_{2}(\mathbf{k})\sigma_y\tau_y+m_{3}(\mathbf{k})\sigma_z\tau_y+m_{4}(\mathbf{k})\sigma_0\tau_x+m_{5}(\mathbf{k})\sigma_x\tau_0\nonumber\\
& +m_{6}(\mathbf{k})\sigma_y\tau_0+m_{7}(\mathbf{k})\sigma_x\tau_z+m_{8}(\mathbf{k})\sigma_y\tau_z+m_{9}(\mathbf{k})\sigma_z\tau_z \text{ .}
\end{align} 

\noindent $m_{3}(\mathbf{k})$, $m_{4}(\mathbf{k})$ and $m_{9}(\mathbf{k})$ are always present (both semimetallic and insulating phases) and are given by
\begin{equation}
m_{3}(\mathbf{k})=\frac{2i}{\omega}\sum\limits_{n=1}^{\infty}\frac{1}{n}\left(M_{n}(\mathbf{k})P_{-n}(\mathbf{k})-M_{-n}(\mathbf{k})P_{n}(\mathbf{k})\right) \text{ ,}
\end{equation}
\begin{equation}
m_{4}(\mathbf{k})=\frac{2i}{\omega}\sum\limits_{n=1}^{\infty}\frac{1}{n}\left(M_{n}(\mathbf{k})Q_{-n}(\mathbf{k})-M_{-n}(\mathbf{k})Q_{n}(\mathbf{k})\right) \text{ ,}
\end{equation}
\begin{equation}
m_{9}(\mathbf{k})=-\frac{2i}{\omega}\sum\limits_{n=1}^{\infty}\frac{1}{n}\left(P_{n}(\mathbf{k})Q_{-n}(\mathbf{k})-P_{-n}(\mathbf{k})Q_{n}(\mathbf{k})\right) \text{ .}
\end{equation}

\noindent On the other hand, $m_{1}(\mathbf{k})$, $m_{6}(\mathbf{k})$ and $m_{7}(\mathbf{k})$ only exist in strained Cd$_3$As$_2$ with $B_{1g}$-type strain. They take the forms 
\begin{equation}
m_{1}(\mathbf{k})=\frac{2i}{\omega}\sum\limits_{n=1}^{\infty}\frac{1}{n}\left(M_{n}(\mathbf{k})B_{-n}(\mathbf{k})-M_{-n}(\mathbf{k})B_{n}(\mathbf{k})\right) \text{ ,}
\label{eq:m1}
\end{equation}
\begin{equation}
m_{6}(\mathbf{k})=-\frac{2i}{\omega}\sum\limits_{n=1}^{\infty}\frac{1}{n}\left(B_{n}(\mathbf{k})P_{-n}(\mathbf{k})-B_{-n}(\mathbf{k})P_{n}(\mathbf{k})\right) \text{ ,}
\end{equation}
\begin{equation}
m_{7}(\mathbf{k})=-\frac{2i}{\omega}\sum\limits_{n=1}^{\infty}\frac{1}{n}\left(B_{n}(\mathbf{k})Q_{-n}(\mathbf{k})-B_{-n}(\mathbf{k})Q_{n}(\mathbf{k})\right) \text{ .}
\label{eq:m7}
\end{equation}

\noindent Lastly, $m_{2}(\mathbf{k})$, $m_{5}(\mathbf{k})$ and $m_{8}(\mathbf{k})$ are only present if strain of type $B_{2g}$ is applied. They have the same form as Eqs.~\eqref{eq:m1}-\eqref{eq:m7}, respectively, but with $M_{n}(\mathbf{k})$, $P_{n}(\mathbf{k})$ and $Q_{n}(\mathbf{k})$ replaced by $\tilde{M}_{n}(\mathbf{k})$, $\tilde{P}_{n}(\mathbf{k})$ and $\tilde{Q}_{n}(\mathbf{k})$.

\section{Surface states}
In this section we explain how we obtained the surface states (SS) shown in the main text for both the semimetallic and the insulating phases of Cd$_3$As$_2$. For the discussion that follows, it is necessary to define two coordinate systems: hereafter, $(x,y,z)$ is the material frame of reference, where the $\hat{x}$, $\hat{y}$ and $\hat{z}$ are parallel to the crystallographic axes as in Sec.I. Since we are dealing with a tetragonal crystal, $\hat{k_x}$, $\hat{k_y}$ and $\hat{k_z}$ are also aligned with $[100]$, $[010]$ and $[001]$, respectively. Besides, $(x',y',z')$ corresponds to the laboratory frame of reference, and $k'_x$, $k'_y$ and $k'_z$ denote the crystal momentum in this coordinate system. As we show in more detail later, $(x,y,z)$ and $(x',y',z')$ are related by a unitary transformation that depends on which material surface we want to probe for SS.     

We calculate the SS by searching for evanescent states at the $z'=0$ termination of a semi-infinite slab of Cd$_3$As$_2$ defined in the $z'<0$ region. Instead of writing a tight-biding model with open (periodic) boundaries along $z'$ ($x'$ and $y'$) and solve for surface states, here we calculate them directly from the $k.p$ model using an approach similar to that from Ref.~\cite{PabloPRB2020}. 
Evanescent states at $z'=0$ have to fulfill two conditions: (i) $\psi_{\mathbf{k}'}(z'=0)=0$ and (ii) $\psi_{\mathbf{k}'}(z'\rightarrow -\infty)=0$, where $\psi_{\mathbf{k}'}(\mathbf{r}')=u_{\mathbf{k}'}(\mathbf{r}')e^{i\mathbf{k}'\cdot\mathbf{r}'}$ are the Block wave-functions and $u_{\mathbf{k}'}(\mathbf{r}')$ are eigenstates of the first-quantized Hamiltonian $\hat{h}(\hat{U}_{(hkl)}^{-1}\mathbf{k}')$. Here, $\hat{h}(\mathbf{k})$ refer to either the bare first-quantized Hamiltonian [Eq.~\eqref{eq:hk}, Eq.~\eqref{eq:hB1g} or Eq.~\eqref{eq:hB2g}] if light is absent, or to the effective first-quantized Floquet Hamiltonian [Eq.~\eqref{eq:heffSI}] if light is taken into account. The matrix $\hat{U}_{(hkl)}$ is the unitary transformation that aligns the normal to crystal surface labeled by the Miller indices $(hkl)$ with $\hat{z}'$. As an example, let's say that we want to calculate the SS at the (112) surface. The normal to such surface points at the direction $[1,1,\frac{1}{2}]=a(\hat{x}+\hat{y})+\frac{c}{2}\hat{z}=a(\hat{x}+\hat{y}+\hat{z})$, and therefore, $\hat{z}'=(\hat{x}+\hat{y}+\hat{z})/\sqrt{3}$. Besides, the two orthogonal direction in the (112), $[1\bar{1}0]=a(\hat{x}-\hat{y})$ and $[11\bar{1}]=a(\hat{x}+\hat{y}-2\hat{z})$, determine the $x'$ and $y'$ axes. This give us the transformation relation between laboratory and material frames of reference, $\mathbf{r'}=\hat{U}_{(112)}\mathbf{r}$ (and, equivalently $\mathbf{k'}=\hat{U}_{(112)}\mathbf{k}$), where
\begin{equation}
   \hat{U}_{(112)}=\begin{pmatrix}
1/\sqrt{2} & -1/\sqrt{2} & 0 \\
1/\sqrt{6} & 1/\sqrt{6} & -2/\sqrt{6}\\
1/\sqrt{3} & 1/\sqrt{3} & 1/\sqrt{3}
\end{pmatrix}  \text{ .}
\end{equation}

\noindent Similarly, for the $(1\bar{1}0)$ considered in the main text, 
\begin{equation}
     \hat{U}_{(1\bar{1}0)}\equiv \begin{pmatrix}
1/\sqrt{2} & 1/\sqrt{2} & 0 \\
0 & 0 & 1 \\
1/\sqrt{2} & -1/\sqrt{2} & 0 
\end{pmatrix} \text{ .}
\end{equation}

Condition (ii) above is satisfied for complex $k'_{z}$ with a negative imaginary part $\kappa<0$. For each point in the surface Brillouin Zone $(k'_x,k'_y)$ and for a fixed energy $E$, we look for the set of complex $k'_z$ that are solutions of the characteristic polynomial  
\begin{equation}
    \text{det}\left(\hat{h}(\hat{U}^{-1}_{hkl}\mathbf{k}')-E\mathbb{1}\right)=0 \text{ ,}
    \label{eq:characteristic}
\end{equation}

\noindent and select only those with $\kappa<0$. Besides, only solutions with $\kappa$ that differ by more than a numerical precision of $10^{-5}$ are chosen. Not all of these complex $k'_z$ solutions gives rise to true eigenstates of the Hamiltonian. Therefore, we substitute each of the solutions back in $\hat{h}(\hat{U}^{-1}_{(hkl)}\mathbf{k}')$ calculate its eigenvalues and eigenvectors. The eigenvectors whose eigenvalue matches the initial energy cut $E$ correspond to localized states that decay exponentially in the region $z'<0$. To obtain the SS, the condition $\psi_{\mathbf{k}'}(z'=0)=0$ still needs to be fulfilled, and this is done by linear combinations of the eigenvectors just collected, which we hereafter denote by
\begin{equation}
\{\left|u_{\mathbf{k}'_{1}}\right\rangle, \left|u_{\mathbf{k}'_{2}}\right\rangle, \cdots\} \text{ .}
\label{eq:set_solutions}
\end{equation} 
\noindent Note that since the coefficients of Eq.~\eqref{eq:characteristic} are real, its solution comes in complex conjugated pairs. Therefore, the maximum number of vectors in the set \eqref{eq:set_solutions} is half of the order of Eq.~\eqref{eq:characteristic}. 

A linear combination of the Block wave-functions, at $z'=0$, constructed with the elements of Eq.~\eqref{eq:set_solutions}
\begin{equation}
    \phi(z'=0)=\sum\limits_{i}\alpha_{i}u_{\mathbf{k}',i}(0) \text{ ,}
    \label{eq:combination}
\end{equation}
\noindent fulfills the condition $\phi(z'=0)=0$ if the determinant of the matrix $M$ whose columns are the $\left|u_{\mathbf{k}'_{i}}\right\rangle$ entering in Eq.~\eqref{eq:combination} is zero. In this case, a SS with energy $E$ exists at $(k'_x,k'_y)$. Small deviations from zero means that the surface state does not exactly vanish at $z'=0$, but its amplitude is small. We accept results states with $|det M|\leq 5\times 10^{-3}$. This analysis is applied to all points a the grid $-0.05 \text{\AA}^{-1}\leq k'_x,k'_y\leq 0.05$\AA$^{-1}$, where the $k.p$ model is valid and these points are colored according to the value of $det(M)$. 

Importantly, since $\left|u_{\mathbf{k}'_{i}}\right\rangle$ is 4-dimentional, we need, at least, four different states in Eq.~\eqref{eq:set_solutions} to be able to get $\phi(z'=0)=0$. This implies that, in the absence of light, since all the bands are two-fold degenerate, a minimum of two complex $k'_{z}$ with $\kappa<0$ needs to be found in Eq.~\eqref{eq:characteristic}. When the light field is included, on the other hand, the degeneracy of the bands is lifted (except at the Weyl nodes) and the minimum number of distinct solutions increases to four. 
%When we find more than four states in Eq.(\ref{eq:set_solutions}), the four of them with larger $|\kappa|$ are selected.

\section{Supplementary figures}

\begin{figure}[H]
    \centering
    \includegraphics[width=0.99\linewidth]{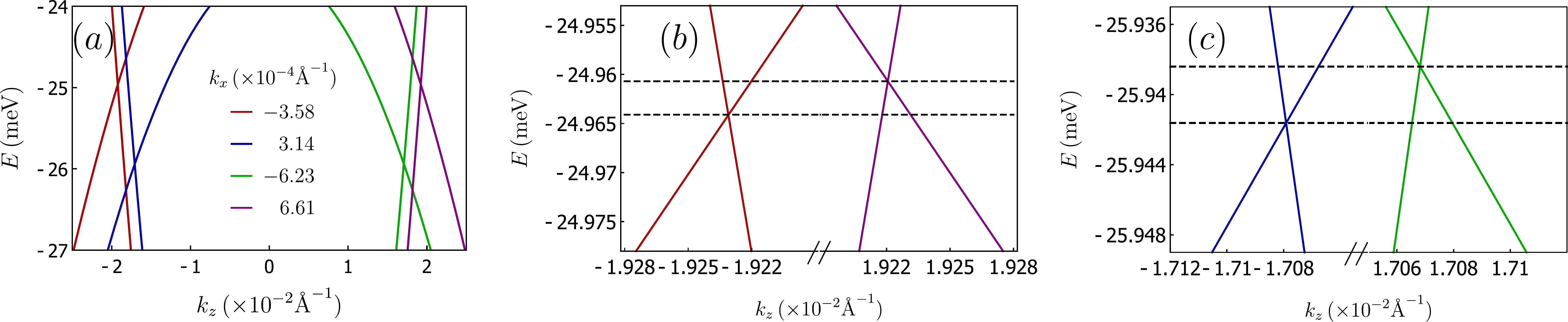}
    \caption{(a) Cuts with $k_x=k_y$ of the bulk bands of Cd$_3$As$_2$ irradiated with BCL. We set the incidence direction normal to (112) surface, $\omega=300$\,meV, $\mathcal{A}_0=2.6\times 10^{-2}$~\AA$^{-1}$ and $\alpha=\pi/2$. Panels (b) and (c) are zooms of panels (a) to highlight the energy separation of the cones.}
\end{figure}

\begin{figure}[H]
    \centering
    \includegraphics[width=0.9\linewidth]{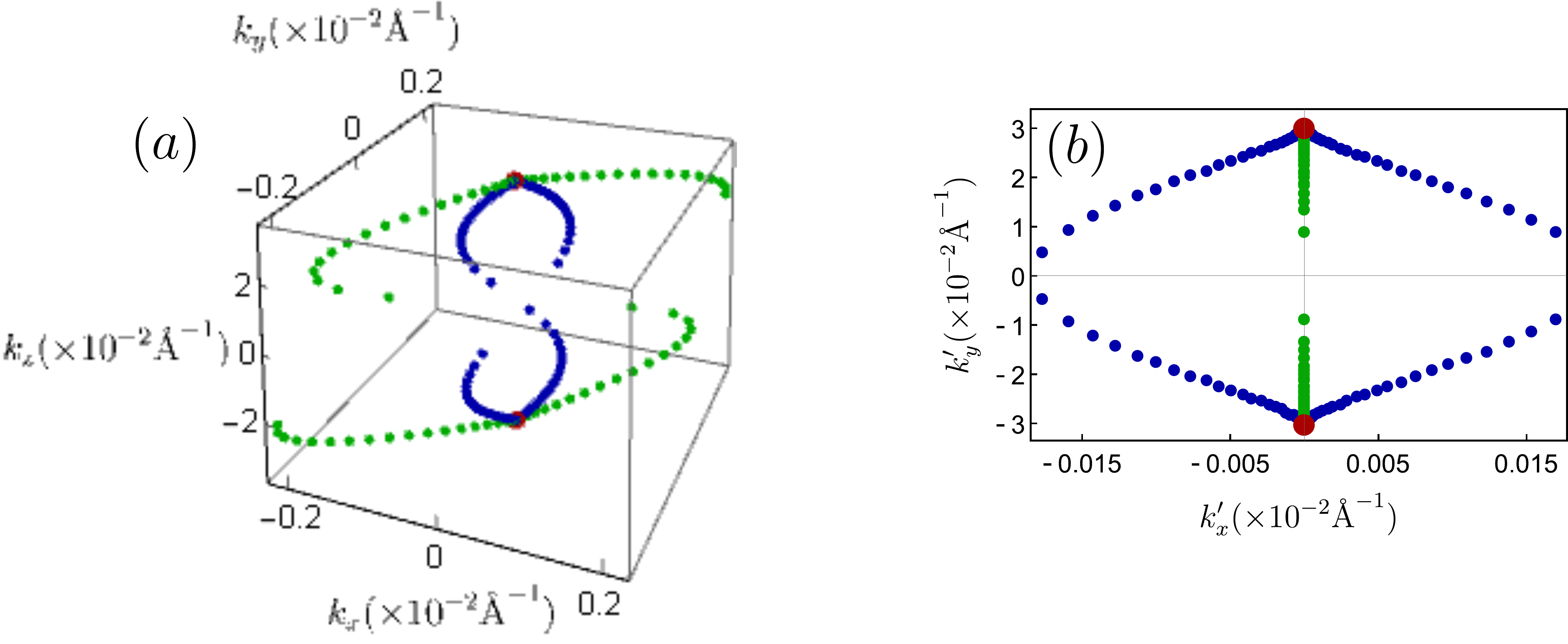}
    \caption{(a) Evolution of the position of the Weyl nodes, in momentum space, as a function of the light when Cd$_3$As$_2$ is irradiated with (i) a CL light (green points) with $0\leq\mathcal{A}_0\leq 3.8\times 10^{-2}$\AA$^{-1}$ and (ii) BCL (blue points) with $0\leq\mathcal{A}_0\leq 2.9\times 10^{-2}$\AA$^{-1}$ and $\alpha=0$. In both cases, we set $\omega=300$~meV. Besides, the red points denote the positions of the original Dirac nodes (in the absence of light). Panel (b) shows the projections of the points of panel (a) in the (112) surface}
\end{figure}

\begin{figure}[H]
    \centering
    \includegraphics[width=0.4\linewidth]{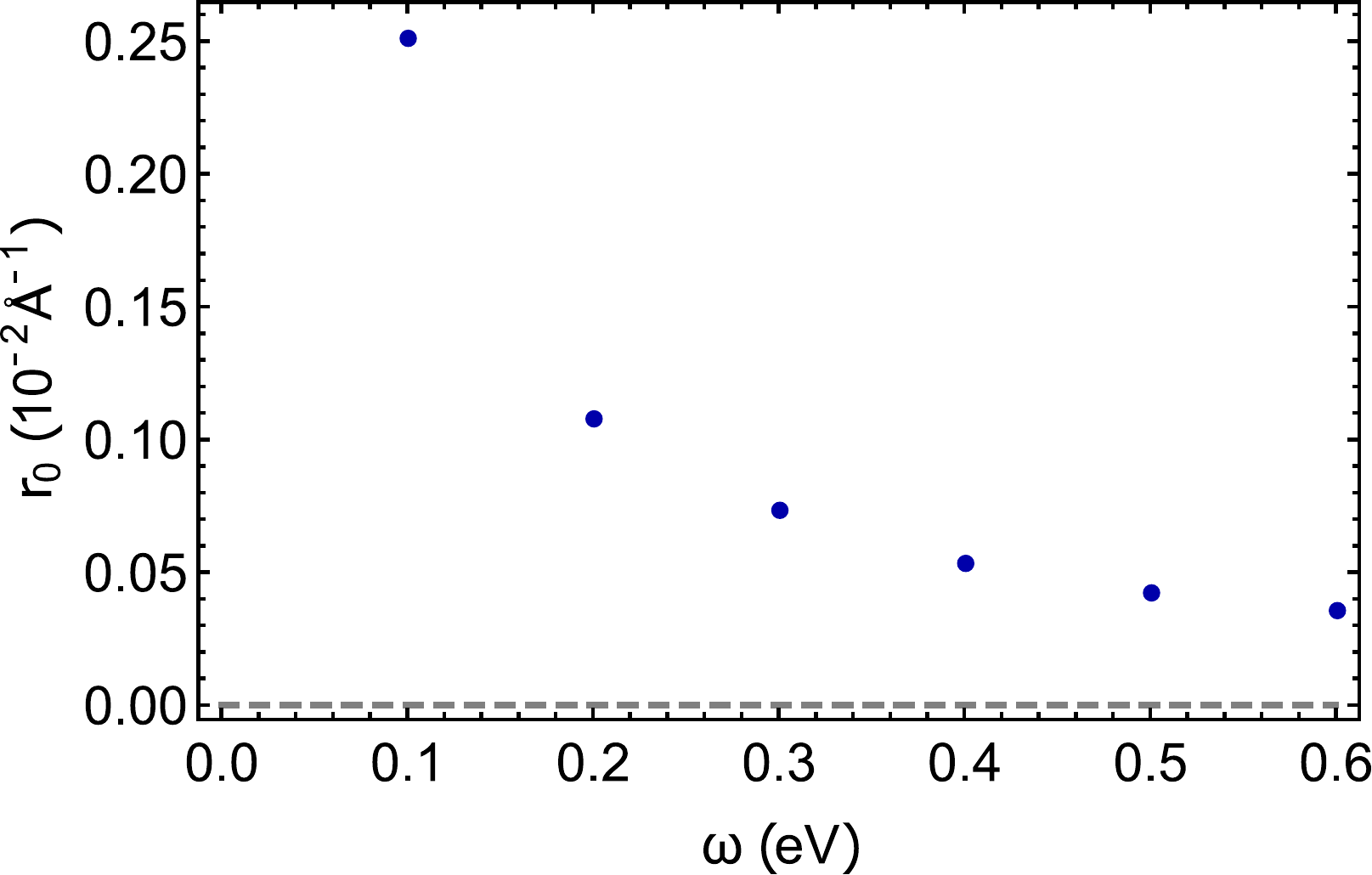}
    \caption{Position $r_0=\sqrt{k_{x}^{'2}+k_{y}^{'2}}$ of the surface Dirac node in the $(1\bar{1}0)$ surface as a function of BCL frequency. We set the incidence direction to be normal to (112) and $\mathcal{A}_0=2.6\times 10^{-2}$~\AA$^{-1}$ and $\alpha=\pi/2$.}
\end{figure}

\begin{figure}[H]
    \centering
    \includegraphics[width=0.55\linewidth]{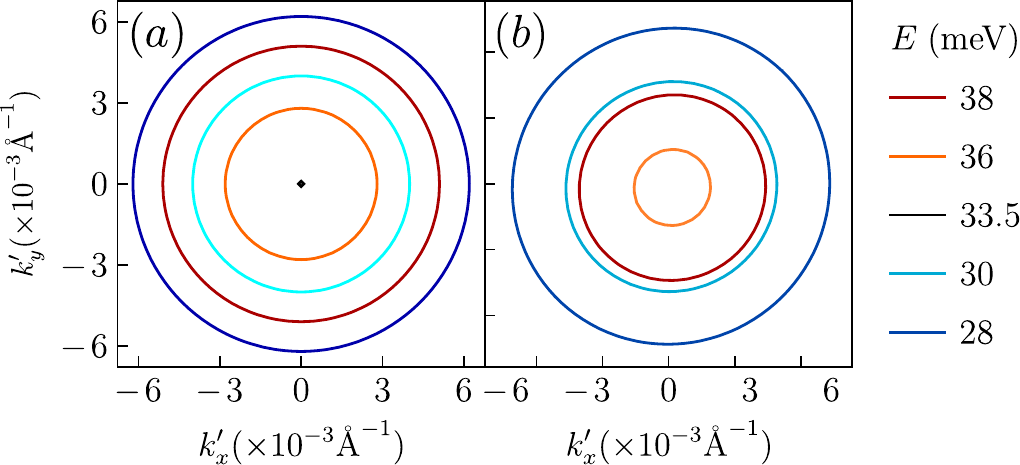}
    \caption{Energy cuts of the surface states of strained Cd$_3$As$_2$ in the (001) surface (a) in the absence of light and (b) in the presence of a CL incident normal to (112). We set the strain type to be $B_{1g}$, $\omega=400$~meV and $\mathcal{A_0}=2.4\times 10^{-2}$\AA$^{-1}$.}
\end{figure}
 %Complete spectrum semimetallic phase
 
 % (Possibly) position of the Weyl nodes for the polarization state used in Fig.2
 
 % Panels (d) and (e) of Fig.3
 
 % Position of the surface Dirac node as function of omega for A_0 fixed
 
 % Position of the surface Dirac node as function  of A_0 for fixed omega

\bibliography{Biblio_BCL}